\documentclass{moriond}

\def\be{\begin{equation}}
\def\ee{\end{equation}}
\def\bea{\begin{eqnarray}}
\def\eea{\end{eqnarray}}

\def\qqbar{$q\bar{q}$}
\def\qqbarq{$q\bar{q}q$}

\def\jpsi{J/\psi}

\def\etal{{\it et al.}}
\def\ie{{\it i.e.}}

\newcommand{\Photo}{}

\begin{document}
\vspace*{4cm}
\title{EXOTIC CHARMONIUM-LIKE STATES AT BESIII}

\author{ KAI ZHU \\ On behalf of BESIII collaboration}

\address{Institute of High Energy Physics, Beijing, 100049, China}

\maketitle\abstract{
The recent measurement results of exotic charmonium-like states, the so called XYZ
particles, at BESIII have been presented. I mainly discussed the charged $Z_c(3900)$ state, its neutral partner, and possible excited states. 
}

\section{Introduction}

Charmonium states are charm quark pair meson $c\bar{c}$, its spectrum
can be calculated by potential model and the theoretical predictions are consistent with
experimental results well. However, at the charmonium states region, recently some new states have been observed, and these so called
XYZ particles cannot be assigned in the charmonium frame easily. On the other
hand, the quantum chromodynamics (QCD) predicts that there are maybe exotic states rather than the usual
baryon ($qqq$) and meson (\qqbar) states. They could be bound gluons
(glueball), \qqbar-pair mixed with excited gluons (hybrids), multi-quark color singlet states
such as: \qqbar\qqbar~(tetra-quark or molecular), \qqbar\qqbarq~(penta-quark),
\qqbar\qqbar\qqbar~(six-quark or baryonium) and \etal. These newly found XYZ states are 
candidates of the exotic states.

BESIII detector is a magnetic spectrometer~\cite{NIM.A614.345} located at BEPCII,
which is a double-ring $e^+ e^-$ collider working at the center-of-mass energy from $2.0$
to $4.6$ GeV. Rich physics potential is at BESIII, that includes light hadron, charmonium,
charm, and R \& QCD physics. The cylindrical
core of the BESIII detector consists of a helium-based main drift chamber (MDC), a plastic
scintillator time-of-flight system (TOF), and a CsI(Tl) electromagnetic calorimeter (EMC),
which are all enclosed in a superconducting solenoidal magnet providing a 1.0 T magnetic
field . The solenoid is supported by an octagonal flux-return yoke with resistive plate
counter muon identifier modules interleaved with steel. The acceptance of charged
particles and photons is 93\% over 4$\pi$ solid angle. The momentum resolution of the
charged particle at 1 GeV/c$^2$ is 0.5\%, and the dE/dx resolution is 6\%. The EMC
measures photon energies with a resolution of 2.5\% (5\%) at 1 GeV in the barrel
(endcaps). The time resolution of TOF is 80 ps in the barrel and 110 ps in the end caps.
Till now, BESIII has collected about $0.6$ B $\psi'$ events, $1.3$ B $J/\psi$ events,
$2.9\ \mathrm{fb}^{-1}$ $\psi(3770)$, and others including scan and continuum data,
\etal. Above $4$ GeV, it has collected about $5\ \mathrm{fb}^{-1}$ data mainly for the studies of XYZ
states.

\section{XYZ particles}
A charged Z state named $Z_c(3900)$ is observed by BESIII~\cite{PhysRevLett.110.252001}
and Belle~\cite{PhysRevLett.110.252002} via process $\pi^\pm (\pi J/\psi)^\mp$, and confirmed with CLEO-c's
data~\cite{PhysLett.B727.366}. The measured mass and width of the three experiments are shown in Table.~\ref{tab:zc}. Analysis shows $Z_c(3900)$ is
strongly coupled to $c\bar{c}$ as well as has electric charge, that indicates it is at least
a $4-$quarks state. Many interpretations are proposed, such as $DD^*$ module, tetra-quark
state, Cusp, and threshold effect, \etal, but none of them is a satisfied explanation yet.

\begin{table}[htp!]
\begin{center}
\caption{Masses and widths of the $Z_c(3900)$ and its neutral partner or excited
  states. The results are default from BESIII without explicit specification. For preliminary results from BESIII, only statistical uncertainties are shown here.}
\label{tab:zc}
\begin{tabular}{l|l|l}
\hline \hline
states & $M(\mathrm(MeV)$ & $\Gamma (\mathrm{MeV})$
\\ \hline
$Z_c(3900)$ (BESIII) & $3899.0 \pm 3.6 \pm 4.9$ & $46 \pm 10 \pm 20$ \\
$Z_c(3900)$ (Belle) & $3894.5 \pm 6.6 \pm 4.5$ & $63 \pm 24 \pm 26$ \\
$Z_c(3900)$ (CLEO-c data) & $3885 \pm 5 \pm 1$ & $34 \pm 12 \pm 4$ \\
$Z_c^0(3900)$  & $3894.8 \pm 2.3$ & $29.6 \pm 8.2$ \\
$Z_c(3885)$ [single tag] & $3883.9 \pm 1.5 \pm 4.2$ & $24.8 \pm 3.3 \pm 11.0$ \\
$Z_c(3885)$ [double tag] & $3884.3 \pm 1.2 $ & $23.8 \pm 2.1 $ \\
$Z_c(4020)$  & $4022.9 \pm 0.8 \pm 2.7$ & $7.9 \pm 2.7 \pm 2.6$ \\
$Z_c^0(4020)$  & $4023.9 \pm 2.2 \pm 3.8$ & fixed to $\Gamma(Z_c(4020))$ \\
$Z_c(4025)$  & $4026.3 \pm 2.6 \pm 3.7$ & $24.8 \pm 5.6 \pm 7.7$ \\
\hline \hline
\end{tabular}
\end{center}
\end{table}

In order to reveal the nature of the $Z_c$ state, BESIII systematically searched its other
decay modes, its charge conjugated partner, and its excited states. Its neutral partner
$Z_c^0(3900)$ has been found via $\pi^0 J/\psi$ spectrum.
%, the shape is displayed in Fig.~\ref{fig:zc0}. 
When BESIII searched another $Z_c$ production mode via $\pi^\pm
(D\bar{D}^*)^\mp$~\cite{PhysRevLett.112.022001} with a single D-tag method, the so called
$Z_c(3885)$ is found. It's mass and width are consistent to the $Z_c(3900)$. This particle
and its decay mode has been confirmed by BESIII with a double D-tag method.
%, the shape has been displayed in Fig.~\ref{fig:dt}. 
At higher mass region, a charged $Z_c(4020)$ is found
via process $\pi^\pm (\pi h_c)^\mp$~\cite{PhysRevLett.111.242001}, as well as its neutral partner
via $\pi^0 (\pi h_c)^0$.
% shown in Fig.~\ref{fig:4020}. 
Close to $Z_c(4020)$, a charged $Z_c(4025)$
is found via process $\pi^\pm (D^*\bar{D}^*)^\mp$~\cite{PhysRevLett.112.132001}. The
masses and widths of these states are listed in Table~\ref{tab:zc} too for comparison. BESIII also
searched $Z_c \to \omega \pi$, however no significant signal has been observed, the upper
limits of production cross sections are $\sigma(e^+ e^- \to Z_c \pi,\ Z_c \to \omega \pi)$
$< 0.27 \ \mathrm{pb}$ and $< 0.18 \ \mathrm{pb}$ at $90\%$ credible level for the center-of-mass energies of $4.23$ GeV
and $4.26$ GeV, respectively.

\begin{figure}[ht!]
 \centering
 \includegraphics[width=0.8\textwidth]{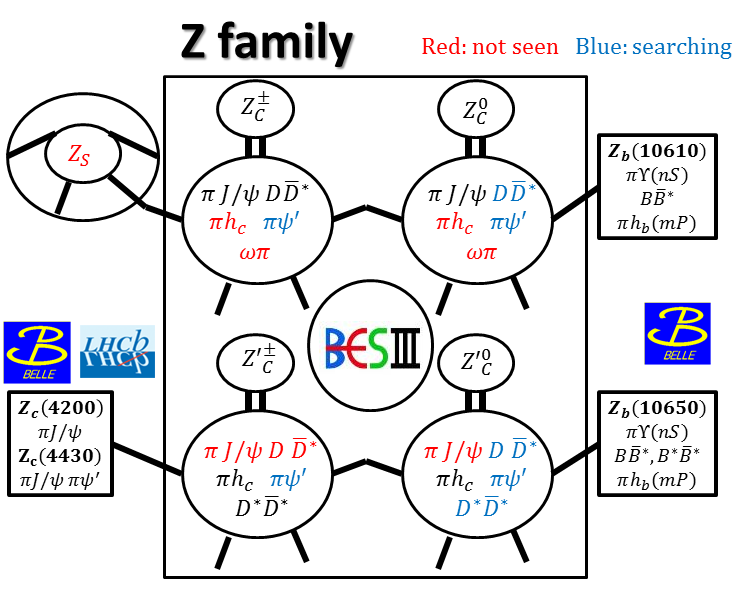}
 \caption{A proposed Z family chart.}
 \label{fig:family}
 \end{figure}

Due to their similar masses and widths, we may assume $Z_c(3900)$ and $Z_c(3885)$ are the
same state $Z_c$, while $Z_c(4020)$ and $Z_c(4025)$ are same as an excited state
$Z'_c$. For even higher $Z_c$ excited state, Belle observed
$Z_c(4200)$~\cite{PhysRev.D90.112009} and $Z_c(4030)$~\cite{PhysRevLett.100.142001} via
$B$ decay ($\pi J/\psi$ and $\pi \psi'$), and $Z_c(4030)$ is confirmed by LHCb via $B^0
\to \psi' \pi K^+$~\cite{PhysRevLett.112.222002}, as well as its $J^P$ is determined as
$1^+$. In order to search $Z_{cs}$, Belle updated its previous $K^+ K^- J/\psi$
measurement to a Dalitz Plot analysis~\cite{PhysRev.D89.072015}, however, no evident structure is
found in $K^\pm J/\psi$ mass distribution under current statistics. Belle has observed
$Z_b(10610)$ and $Z_b(10650)$ in $\pi^+ \pi^- \Upsilon(nS)$~\cite{PhysRevLett.108.122001}
\cite{arxiv:1105.4583}, $\pi^+ \pi^- h_b(mP)$~\cite{PhysRevLett.108.032001} and
$[B\bar{B}^*]^\pm \pi^\mp$~\cite{arxiv:1209.6450} final states. Due to above information, a $Z$ family
chart is  proposed in Fig.~\ref{fig:family}.

$X(3872)$ was first observed via $B\to K(\pi^+ \pi^- J/\psi)$ process by
Belle~\cite{PhysRevLett.91.262001}, and its mass is close to the $D^0 D^{*0}$ threshold and
its width very narrow. Its $J^{PC}$  has been determined by CDF~\cite{PhysRevLett.98.132002} and
LHCb~\cite{PhysRevLett.110.222001} to be $1^{++}$. The partial width
measurement shows that it takes a $50\%$ chance to decay via open-charm channel and at the
$O(\%)$ via charmonium. Since its discovery, many interpretations of its nature have been proposed, such as $D^0 D^{*0}$ module, hybrid, $\chi_{c1}(2P)$, 
and tetra-quark states, \etal. However no one is very satisfactory and more studies are needed to understand its nature
. Recently BESIII reported a new production mode of $e^+ e^- \to \gamma (\pi^+
\pi^- J/\psi)$~\cite{PhysRevLett.112.092001}, with data samples collected at center-of-mass energies from $4.009$ to $4.420~\mathrm{GeV}$. Before BESIII's discovery, only in the pp collision and B decays $X(3872)$ has been found.
 
$X(3823)$ was firstly seen by Belle in $B\to \chi_{c1} \gamma K$ decays as a narrow peak
in the invariant mass distribution of the $\chi_{c1} \gamma$
system~\cite{PhysRevLett.111.032001}. Its properties are consistent with the $\psi_2(1^3D_2)$
charmonium state. At BESIII, $X(3823)$ is observed via $e^+ e^- \to \pi^+ \pi^- X(3823)
\to \pi^+ \pi^- \gamma \chi_{c1}$, and there is evidence that it is may from $Y(4260)$ decay.
%, as shown in Fig.~\ref{fig:x3823}. 

In last decade, lots of Y states have been observed, for example $Y(4008)$, $Y(4260)$,
$Y(4360)$, $Y(4630)$, and $Y(4660)$, \etal. They are produced via initial state radiative
(ISR) processes in $e^+ e^-$ collision, that means they are vector states. However, the
total number of these Y states is absolutely larger than that of the theoretical prediction by
potential model for charmonium around this energy region. One thing is interesting that these Y
states are only found in $\pi^+ \pi^-
J/\psi$~\cite{PhysRevLett.95.142001,PhysRevLett.99.182004}, $\pi^+\pi^-
\psi(2S)$~\cite{PhysRevLett.99.142002}, and $\Lambda^+_c
\Lambda^-_c$~\cite{PhysRevLett.101.172001} final states, but no evidence from open charm, \ie $Y\to
D^{(*)}D^{(*)}$~\cite{PhysRev.D77.011103,PhysRevLett.98.092001,PhysRevLett.100.062001}.

At BESIII, abundant structures are observed in the cross section shapes from $4.0$ to
$4.6$ GeV via different processes, that are from $e^+ e^-$ to $\pi^+ \pi^-
h_c$~\cite{PhysRevLett.111.242001}, $\omega \chi_{c0}$~\cite{PhysRevLett.114.092003}, $\eta
 \jpsi$,  $\pi^0 \pi^0 \jpsi$, $\jpsi \eta \pi^0$, $\pi^+ \pi^- \gamma \chi_{c1}$, $\eta'
 \jpsi$, $\gamma \phi \jpsi$ and $\gamma \chi_{cJ}$. But the statistics are limited, then larger data samples are necessary and possible interferences need be considered to understand them correctly.

\section{Summary}
A relatively systematic study on exotic charmonium-like states, \ie XYZ particles, has been
performed at BESIII. Abundant and interesting results are obtained, and that has greatly improved our knowledge about these exotic states.  However, the natures of
the these exotic states are still mysterious, the relations between them are unclear, and
some expected states and decay modes are missing. To shed light on the puzzles, more
decay channels should be studied carefully, partial wave analysis (PWA) is needed to
clarify their quantum number, \ie\ $J^{JC}$, and a fine scan from $3.8$ to $4.6$ GeV at
BESIII may be helpful. More results from BESIII will come soon with analyses are under way.

\section*{Acknowledgments}
I am grateful to the Moriond2015 committee for the organization of this
very nice conference.

\end{document}